\documentclass[preprint,12pt]{elsarticle}
\usepackage{amssymb}
\usepackage{url}
\usepackage{graphicx}
\usepackage{mathabx}
\usepackage[usenames,dvipsnames]{xcolor}
\usepackage[section]{placeins}
\usepackage[normalem]{ulem}
\usepackage{xcolor}
\headsep=-.cm
\newcommand\crule[3][black]{\textcolor{#1}{\rule{#2}{#3}}}

\colorlet{darkgreen}{green!50!black}
\colorlet{brightyellow}{yellow!75!red}
\colorlet{orange}{red!50!yellow}
\colorlet{darkred}{red!80!black}
\colorlet{darkblue}{blue!50!black}

\usepackage{soul}

\def\rhovec{\mbox{\boldmath $\rho$}}

\newcommand{\beq}{\begin{eqnarray}}
\newcommand{\eeq}{\end{eqnarray}}

\journal{Physics Letters B}

\begin{document}

\begin{frontmatter}
\title{$^7$H ground state as a {$^3$H+4n} resonance
\author{Emiko Hiyama}           \address{Department of Physics, Tohoku University, Sendai, 980-8578, Japan and \\  RIKEN Nishina Center, 2-1 Hirosawa, Wako 351-0106, Japan}
\author{Rimantas Lazauskas} \address{IPHC, CNRS/IN2P3, Universit\'e de Strasbourg, 67037 Strasbourg, France}
\author{Jaume Carbonell}       \address{Universit\'e Paris-Saclay, CNRS/IN2P3, IJCLab, 91405 Orsay, France}
}

\begin{abstract}
We have investigated the possible existence of  a $^7$H resonant state, considered  as a five-body system 
consisting of a $^3$H  core with four valence neutrons. 
To this aim, an effective n-$^3$H potential is constructed in order to
reproduce the low energy elastic neutron scattering on $^3$H  phase
shifts and the $^5$H resonant ground state in terms of $^3$H-n-n
system. The variational Gaussian Expansion Method is used to solve the 5-body Schr\"{o}dinger equation, while the resonant
state parameters were estimated by means of the stabilization method. 
We have not found any sign of a narrow low energy resonance in the vicinity of $^3$H+4n threshold. 
However,  we have identified  a very broad structure at $E_R\approx 9$ MeV above this threshold, which corresponds to the $^7$H J$^{\pi}$=1/2$^+$  ground state.
In the vicinity of this state, we have also identified a broad structure corresponding to the ground state of $^6$H isotope  with quantum numbers $J^{\pi}=2^-$.  
 \end{abstract}


\end{frontmatter}

%\tableofcontents

\section{Introduction} \label{intro}

Aside from the well-known  stable isotopes - deuteron $^2$H (d) and tritium $^3$H  (t) - the H isotopic chain  extends beyond
these two familiar nuclei with well-defined resonant states in $^4$H  --  clearly seen in the low energy n-t  cross section-- and $^5$H  system
\cite{Young_PR173_1968,Korsheninnikov_PRL87_2001,Golovkov_et_al_2003_04_05,Wuosmaa_PRC95_2017}.
The possibility for the simplest nucleus, a single proton, to form an isotopic  state as exotic as $^7$H constitutes an exciting
challenge both from  experimental as well as from  theoretical points of view.

The first indication of a resonant $^7$H state came in 2003 from RIKEN \cite{7H_RIKEN_Korsheninnikov_PRL90_2003} study of p($^8$He,2p)$^7$H reaction.  These authors conclude  the
existence of a $^7$H resonance near the  t+4n threshold, although they were not able to determine the resonance parameters.

This was immediately followed by a series of experiments in GANIL
using a $^8$He beam of $\approx$ 15 A MeV. The first  one (E465S)
\cite{7H_GANIL_Fortier_AIPCP912_2007,7H_GANIL_Baumel_ISPUN0_2007}
was based on the transfer reaction  d($^8$He,$^3$He)$^7$H. 
They measured  a clear structure in the missing mass spectrum, which
was attributed to the presence of a  resonant state with $E_R$=1.56 $\pm$ 0.27 above the  t+4n threshold and  $\Gamma$=1.74$\pm$ 0.72 MeV.

In the next GANIL experiment (E406S) \cite{7H_GANIL_PRL99_2007,7H_GANIL_PRC78_2008}, 
the $^{12}$C($^8$He,$^{13}$N)$^7$H transfer reaction was studied by employing a $^8$He beam of 15.4 A MeV and a $^{12}$C gas target.
Regardless of the  small statistical evidence (7 events),  these authors suggested the existence of a very narrow resonant $^7$H state at
$E_R$=0.57+0.42-0.21 MeV and an estimated width  $\Gamma$=0.09+0.94-0.06 MeV.
The same result has been recently published in \cite{7H_GANILbis_PLB829_2022}.  				

GANIL results were not fully affirmed in a 2010 RIKEN experiment \cite{7H_RIKEN_PRC81_2010}. 
Nevertheless, a well-pronounced peak was observed and associated with a $^7$H  state, resulting in a peculiar structure at $\sim$2 MeV above the t+4n threshold.

A recent experiment \cite{7H_Dubna_PRL124_2020} at the Flerov Laboratory of Nuclear Reactions (JINR, Dubna) %at the ACCULINNA-2
employed a $^8$He beam  of 26 A MeV to investigate the  $^2$H($^8$He,$^3$He)$^7$H reaction and suggested the existence of a
$^7$H ground state at E=2.0(5) MeV as well as the first excited state at E$^*$=6.5(5) MeV with $\Gamma$=2.0(5) MeV.

We should also mention the unsuccessful attempt of Ref. \cite{Dubna_PLB588_2004} to observe a long-living quasi stable
$^7$H nucleus in the reaction $^2$H($^8$He,$^7$H)$^3$He. The authors  of a former work estimated the $^7$H lifetime to be less
than 1 ns and a lower limit of 50-100 keV for the $^7$H energy above the t + 4n breakup threshold.

\bigskip
From the theoretical side,  little is known despite the growing interest in exotic nuclear systems. One needs to clarify the
tendency -- for or against the stability -- of the H isotopic chain when moving from p+3n to p+4n and to p+6n. Our previous {\it ab-initio} calculation  on $^4$H and $^5$H
\cite{5H_LHC_PLB791_2019} shows a gain of stability (i.e. proximity to the E-real axis) in the computed S-matrix poles when
moving from $^4$H to $^5$H ground state (see Fig 4 of this reference). If the results of the GANIL second experiment
\cite{7H_GANIL_PRL99_2007,7H_GANIL_PRC78_2008} would be confirmed, they will indicate a further increase  when moving to p+6n.
%,\edit{and would make more reliable the eventual  observation of 6 or 8 multineutron resonant states.}

However, only a few attempts have been made in this direction. 
In Ref. \cite{Natasha_PRC65_2002} the energy of $^7$H was roughly estimated  to  E($^7$H)$\sim$1.4 MeV (above t+4n) by  using a purely S-wave Volkov  
potential \cite{Volkov_NP74_1965} and an exponential extrapolation of energies calculated up to the Hyperspherical quantum number K= K$_{min}$+6.

Using a variant of the Volkov potential, one could also estimate the energy of $^7$H in \cite{7H_RIKEN_Korsheninnikov_PRL90_2003}.
Since the experimental $^8$He binding energy is 31.4 MeV, we can
expect the binding energy  of seven-body $^7$H   to be -5.4 MeV, which is about 3 MeV above the t+4n threshold.

A more realistic estimation was realized in \cite{Aoyama_Itagaki_NPA738_2004} based on the same Volkov
interaction. In order to solve the 7-body problem, these authors combined the Antisymmetrized Molecular Dynamics (AMD)
approach with generator coordinate and stochastic variational methods. They found a state at E($^7$H)$\sim$ 7 MeV above the t+4n
threshold. All the calculations were performed within the bound state approximation, failing to estimate the width of a such high energy state. 
Some years later, these authors \cite{Aoyama_Itagaki_PRC80R_2009} realized a more advanced AMD
calculation, which suggested the presence of a state at E($^7$H) $\sim$ 4 MeV above the  t+4n threshold,  
again without giving its decay width.

Finally, it is worth noticing the  recent  work \cite{Li2_Michel_Zuo_PRC104L06_2021} in which $^7$H was considered
as a state formed from a rigid  $^3$H core and four valence neutrons.
The unbound $^{4-7}$H isotopes have been computed by using No-core
Gamow Shell Model (NCGSM) techniques with different phenomenological
NN interactions. These authors concluded that $^7$H should be a very narrow resonant state, even  the sharpest neutron resonance observed in the light nuclei chart. 
The value of $E_R$, the real part of the complex energy ground state, varied from 2-3 MeV depending on the NN model, and in all cases $\Gamma\approx$ 0.1 MeV. This is a
quite surprising result, since the predicted resonance is found to be well above the $t+4n$ threshold, above the resonant $^4$H+3n and $^5$H+2n
decay channels and, nevertheless, its width is predicted to be much smaller than those  of $^4$H and $^5$H resonant states.

Recently, an experimental  group  at RIBF (NP1512-SAMURAI134) have performed a study aiming to observe the $^7$H resonant states.
The analysis of the collected data is still in progress. Considering this situation, it is timely to study the structure of $^7$H.

Contrary to our previous {\it ab-initio} works on 3n \cite{LC_3n_PRC72_2005},  4n \cite{LC_4n_PRC72_2005,HLCK_PRC93_2016,CLHK_FBS_2017,LHC_PTEP7_2017}, 
$^4$H and $^5$H \cite{5H_LHC_PLB791_2019} we are not able to consider the  $^7$H  system as a genuine 7-N problem. 
Our present work is thus an attempt to compute the $^7$H ground state properties in the 5-body  cluster approximation (t-n-n-n-n),
 which anyway should be the dominant decay channel for  a low energy resonant state.
To this aim, we have  first constructed a n-t local interaction that we have adjusted in order to reproduce the n-t phase shifts.
We have also adjusted a simple t-n-n three-body force to reproduce  the  ground state properties of $^5$H, now described as a t-n-n system.  
It should be noted, however, that the experimentally determined energy and decay widths of $^5$H represent only an effective response to a particular reaction mechanism
 leading to the short-living  formation and instantaneous decay of a  $^5$H state. 
Its value is thus  strongly dependent of the particular reaction mechanism. 
Therefore, the t-n-n three-body force used in our  model  was adjusted in  order to reproduce the $^5$H resonance parameters computed in our previous {\it ab initio} calculation \cite{5H_LHC_PLB791_2019}.

The paper is organized as follows.  Section \ref{Formalism} contains the formal aspects of the problem. These are essentially
the t-n potential (\ref{Vtn}), the Gaussian Expansion Method (GEM) used to solve the five-body Schr\"{o}dinger equation for the
t-n-n-n-n system (\ref{GEM}) and the Stabilization Method (SM) used to  estimate  the
complex energies  of resonant states (\ref{SM}). Our numerical results for $^6$H and $^7$H  resonant state are given in Section
\ref{Results} and Section \ref{Conclusion} contains some concluding remarks.

%%%%%%%%%%%%%%%%%%%%%%%%%%%%
\section{Formalism}\label{Formalism}

An {\it ab initio} 7-body calculations of a resonant state is beyond the existing technical capabilities. Since the ground state
of $^7$H decays essentially into the five-body $^3$H+4n channel, we have decided to describe its ground state as a 5-body system, based on a solid  tritium core interacting with four valence nucleons. 

At the current stage,   our 5-body code based on Gaussian expansion method is not able to handle tensor interactions, and therefore our t-n interaction model had to omit such terms. 
On the one hand, it is well known that the tensor force is crucial in order to describe  quantitatively the few-nucleon data. 
On the other hand, light nuclei fulfill quite well  the criteria of Effective Field Theory
\cite{UvK_NPA645_273,Ebra_PRP428_2006,PNa_RPP80_2017,MGA_PRC100_2019},
according to which  the main features he weakly bound nuclei rely on very few {parameters} of the NN interaction. 
Our previous studies of the lightest hydrogen isotopes \cite{5H_LHC_PLB791_2019} clearly confirm this point. 
Indeed, by considering the phenomenological MT I-III NN potential, which is limited to
S-wave, we were able to  qualitatively describe the n-t scattering phase shifts and the position of $^5$H resonant state. 
We have therefore decided to build our n-t effective interaction by adjusting its parameters  to reproduce the  t-n phase shifts
calculated by solving the {\it ab-initio} four-nucleon scattering problem with MT I-III.

%%%%%%%%%%%%%%%%%%%%%
\subsection{The n-t potential}\label{Vtn}

\bigskip
For  numerical convenience, we have constructed a  n-t potential in terms of Gaussian regulators
\begin{equation}\label{Vnt}
V^{SL}_{nt} (r) = \delta_{L0} \mid \phi_0 > \lambda_{\infty} < \phi_0\mid   \;+\;  \sum_{i=1}^3  \left[ V_i^c \;+\; (-)^LV_i^P \;+\;  {\hat{S}^2\over 2} V_i^S \;+\; (-)^L {\hat{S}^2\over 2} V_i^{SP}   \right] \; e^{-a_i r^2}
\end{equation}
where $V_i^c$, $V_i^P$, $V_i^S$ and $V_i^{SP} $  are adjustable parameters, and
\[ <\vec{r}\mid \phi_0 > \equiv \phi_0(r)= e^{-a_0r^2} \]
is an infinitely repulsive S-wave contribution accounting for the Pauli forbidden state. In principle, $ \lambda_{\infty}$ should be infinite, however 
we found that  value $\lambda_{\infty}$=10$^5$ MeV is sufficient to ensure an accuracy of 100 keV  for the calculated $^7$H binding energies.
Potential (\ref{Vnt})  is spin and angular momentum dependent but it has neither spin-orbit nor tensor force. 
In spirit, it is rather similar to the  MT I-III interaction used in out previous studies of $^4$H and $^5$H systems \cite{5H_LHC_PLB791_2019}, and will in particular provide
degenerate  $J^{\pi}=0,1,2^-$ triplet P-wave resonant states states in $^4$H (see Table I from \cite{5H_LHC_PLB791_2019}).

\begin{table}[h!]
\begin{center}
\begin{tabular}{ l  l  r  r  r }\\  \hline
i                  &           &   1                 &        2 &   3  \\ \hline
 $\alpha_i$  & fm$^{-2}$      &    0.04458     &  0.271           &     0.1700   \\ \hline
$V^C_i$     & MeV   &  2.4975        &  96.245            &  -57.19  \\
$V^P_i$     & MeV  &   2.4975        &  -96.245           &  57.19 \\
$V^S_i$     & MeV  & -1.00725       &   -13.755          &  0  \\
$V^{SP}_i$ & MeV & -1.00725       &   13.755          &  0  \\ \hline
\end{tabular}
\end{center}
\caption{Parameters of the V$_{nt}$ effective potential (\ref{Vnt}).}\label{Table_Models}
\end{table}

\begin{figure}[h!]
\centering\includegraphics[width=10.cm]{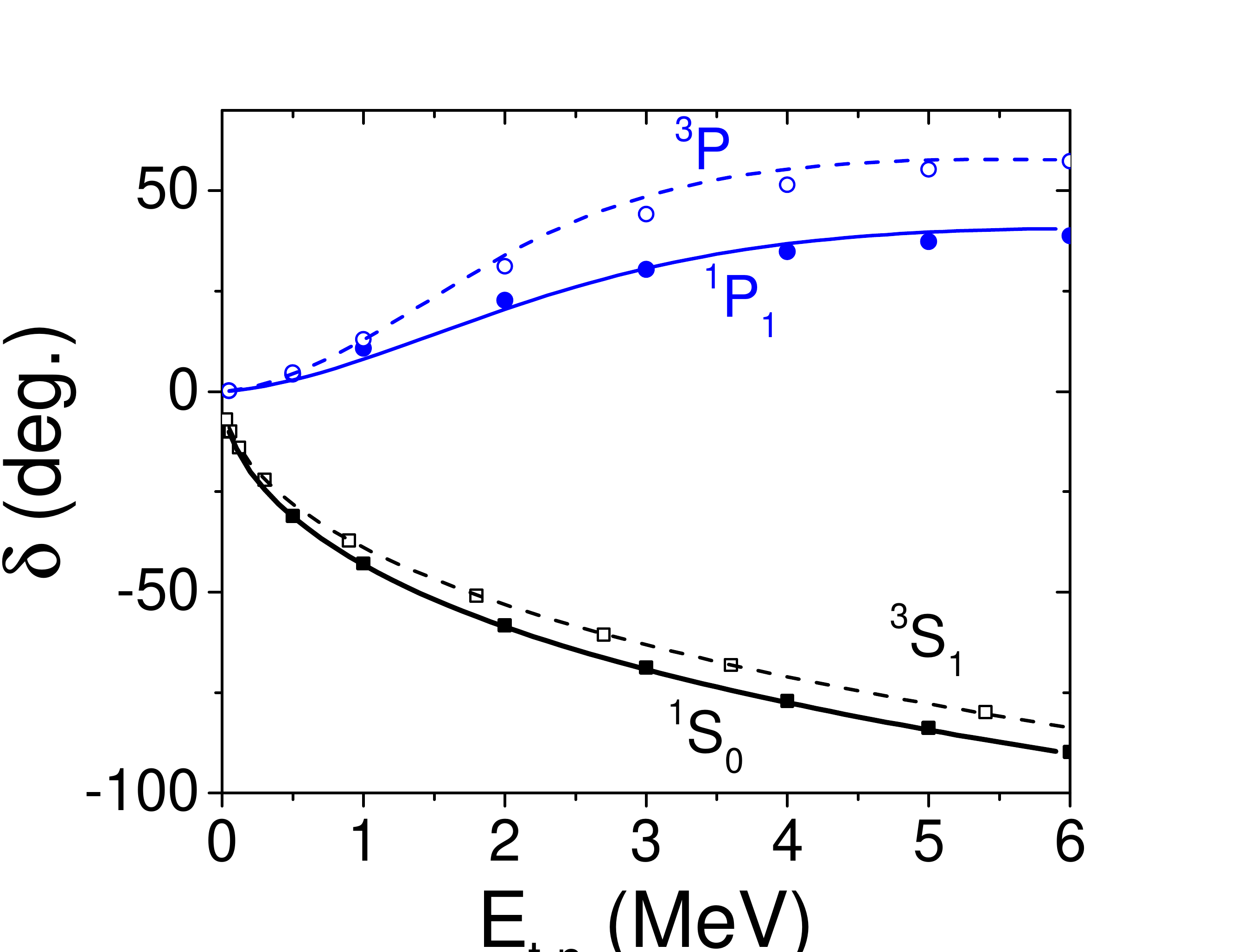} 
\caption{The S- and P-wave n-t phase shifts (solid lines for spin singlet states and dashed lines for triplet ones) computed with the two-body potential (\ref{Vnt}) are 
compared with the results obtained by solving four-nucleon problem with MT I-III potential (cercles and squares)}\label{fig:Phnt}
\end{figure}

The corresponding parameters, adjusted to reproduce the experimental n-t phase shifts,   are provided in Table
\ref{Table_Models} with $a_0$=0.1979 068  fm$^{-2}$. The  results for the n-t phase shifts  obtained  with this
potential (and with ${\hbar^2\over 2\mu_{nt}}$=27.647 MeV fm$^{-2}$) are plotted in Figure \ref{fig:Phnt}. 

As expected, this potential does not support any bound state, however it  has two
broad P-wave resonant states:  one in spin-singlet (S=0) and one in spin-triplet (S=1) channels. 
The corresponding complex energies of these resonant states are given in central column of  Table \ref{Table_Resonance}.
They are in qualitative agreement with the values (right column) obtained by solving
the four-nucleon problem with the  MT I-III potential and the Jost function/boundary condition method to determine the S-matrix pole positions. 

If the agreement between the n-t  two-body model and four-body calculations is not perfect, is essentially due to the fact that the
n-t phase shifts displayed in Figure \ref{fig:Phnt}  are not perfectly reproduced by our two-body  potential (\ref{Vnt}).  
Indeed we had to make some compromises when trying to fit simultaneously all the partial wave  using the simple operator form  (\ref{Vnt}). 
Moreover, when performing this fit we kept in mind  to simultaneously reproduce within our two-body model  the resonant
state of $^5$H and an effort  was made to minimize the effect of t-n-n three-body force. 
On the other hand, the four-nucleon calculations by realistic models predict a $J^\pi=2^-$ resonant
state at slightly higher energy (E$ \approx$1.2-2i) than MT I-III, what justifies the slight discrepancies with our 2-body model prediction

\begin{table}[h!]
\begin{center}
\begin{tabular}{ l  l  r  r }\\ \hline
                         &   V$_{nt}$   (\ref{Vnt})    &   pnnn   \cite{5H_LHC_PLB791_2019}        \\  \hline
$L=1^-,S=0$     &  1.28-2.61 i     & 0.88(5)-2.20(5) i       \\
$L=1^-,S=1$     &  1.33-1.84 i     & 1.08(3)-2.03(3) i          \\  \hline
\end{tabular}
\end{center}
\caption{Positions of $^4$H resonant states (central column)  calculated with the effective 2-body
n-t potential  (\ref{Vnt})   and  (right column) by solving the 4N
problem with the MT I-III interaction. The parameters of (\ref{Vnt}) were adjusted to the n-t scattering phase shifts displayed in Fig.  (\ref{fig:Phnt}).}\label{Table_Resonance}
\end{table}

When describing the $^5$H and $^7$H systems, the n-t interaction has been completed with a neutron-neutron potential. We have taken
the Minnesota  model \cite{NN_Minnesota_1977} which consists in a superposition of two Gaussians with different ranges and provides
the neutron-neutron  low energy parameters  $a_{nn}$=-16.90  fm and $r_{nn}$=2.88 fm.

As it was found in previous studies \cite{DGFJ_NPA786_2007,HOKY_NPA908_2013}, we were not able to
reproduce the experimental  $^5$H ground state (J$^{\pi}$=1/2$^+$) resonance parameters when using the local $V_{nt}$ and $V_{nn}$
interactions, although we were explicitly trying to favour  the appearance of $^5$H resonant state when adjusting parameters of $V_{nt}$.
We have thus adjusted the position of the  $^5$H ground state  by adding a t-n-n
3-body force, having the same form as in Refs. \cite{DGFJ_NPA786_2007,HOKY_NPA908_2013}:

\begin{figure}[h!]
\centering\includegraphics[width=4.cm]{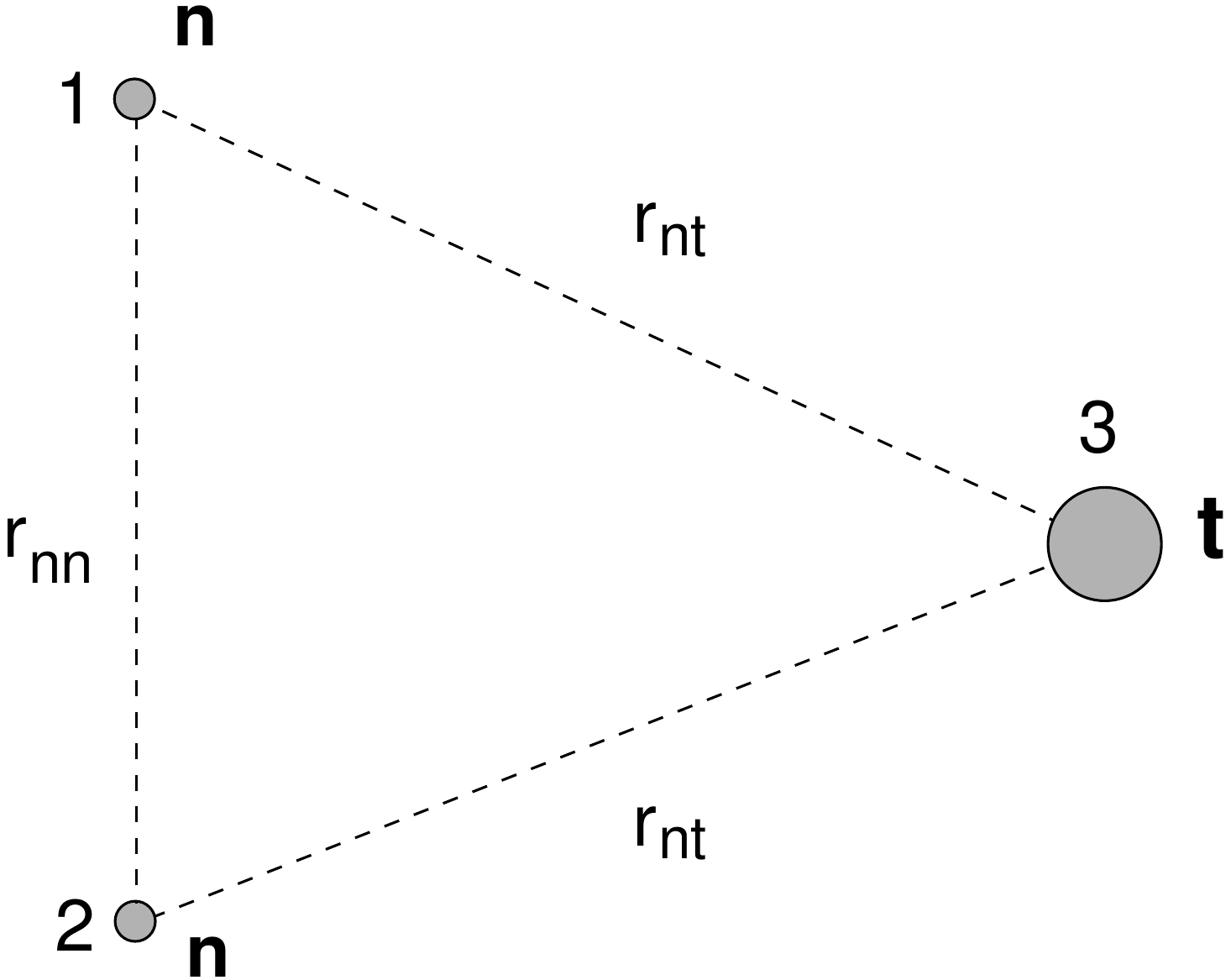}
\caption{$^5$H as a t-n-n three-body system.}\label{FIG_tnn}
\vspace{0.cm}
\end{figure}

\begin{equation}\label{Vtnn}
V_{tnn} (\rho)= -V_0  \; e^{- {\rho^2 \over b_3^2}}
\end{equation}
where
\begin{equation}\label{rho_G}
   \rho^2 = \sum_{i<j} {m_im_j\over m M} \;  (\vec{r}_j-\vec{r}_i)^2     \qquad  M=\sum_im_i
\end{equation}
and $m$ and arbitrary mass. Setting $m=m_n$ one has
\[ \rho^2 =    {m_n \over M} r_{nn}^2 + {m_t \over M} r_{nt}^2 + {m_t \over M} r_{nt}^2     \qquad  M=2m_n+m_t \]
in which $r_{nn}, r_{nt},r_{nt}$ denote the corresponding relative distances, as shown in figure (\ref{FIG_tnn})
\footnote{Notice that in terms of Jacobi coordinates
\begin{eqnarray*}
\vec{x}_3 &=& \sqrt{{1\over m} {2m_1m_2\over (m_1+m_2)}   } \;
(r_2 - r_1)  \cr \vec{y}_3 &=& \sqrt{2m_3(m_1+m_2)\over mM} \;
\left( r_3 - {  m_1 r_1+ m_2 r_2\over m_1+m_2}   \right)
\end{eqnarray*}
on has
\[ \rho_J^2=x_3^2+y_3^2  = 2\rho^2\]}

%%%%%%%%%%%%%%%%%%%%%%%%%%%%%%%%%%%%%%%%%%%%%%
\subsection{Solution of the 3- and 5-body problems with Gaussian expansion method}\label{GEM}

\bigskip

\begin{figure}[h!]
\centering\includegraphics[width=8.cm]{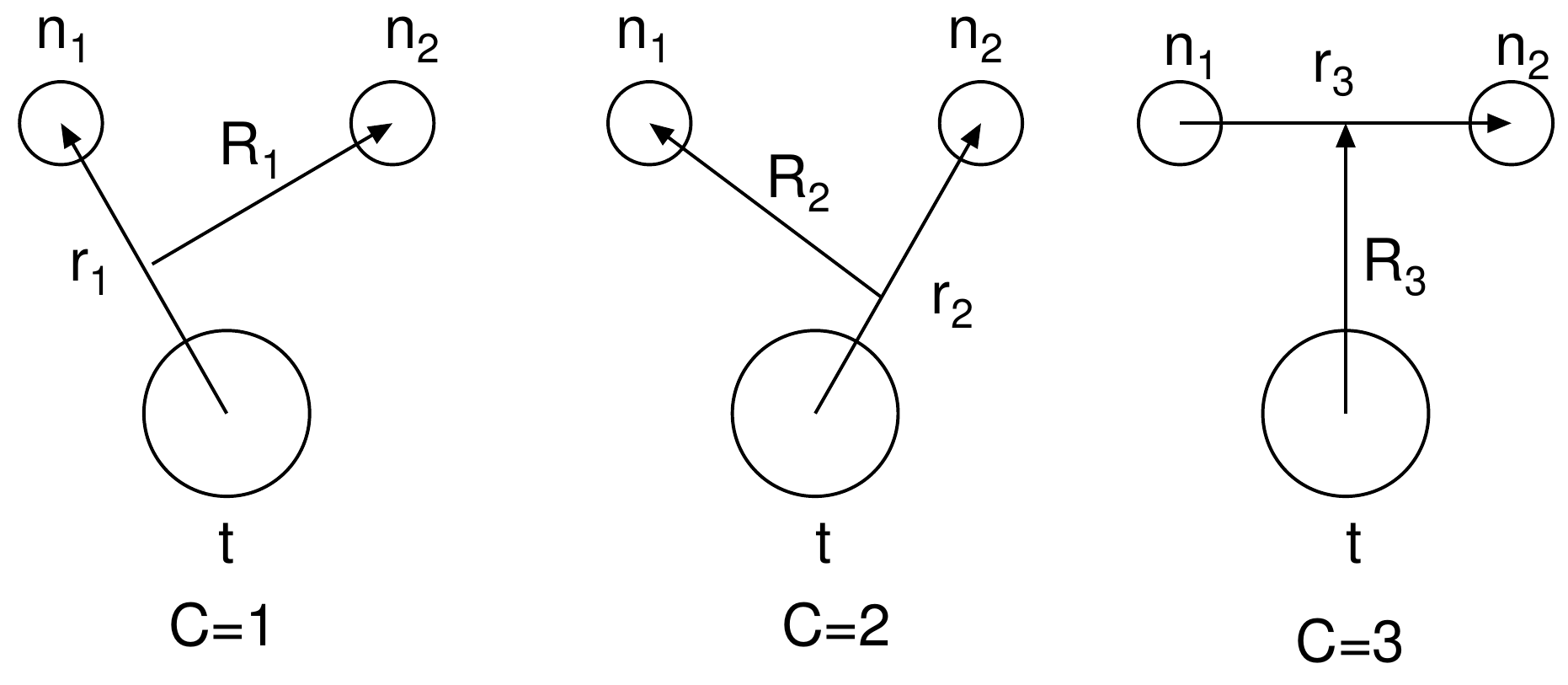}
\caption{Faddeev components and corresponding Jacobi coordinates of the t-n-n sytem.}\label{FC_5H}
\vspace{0.cm}
\end{figure}

\begin{figure}[h!]
\centering\includegraphics[width=10.cm]{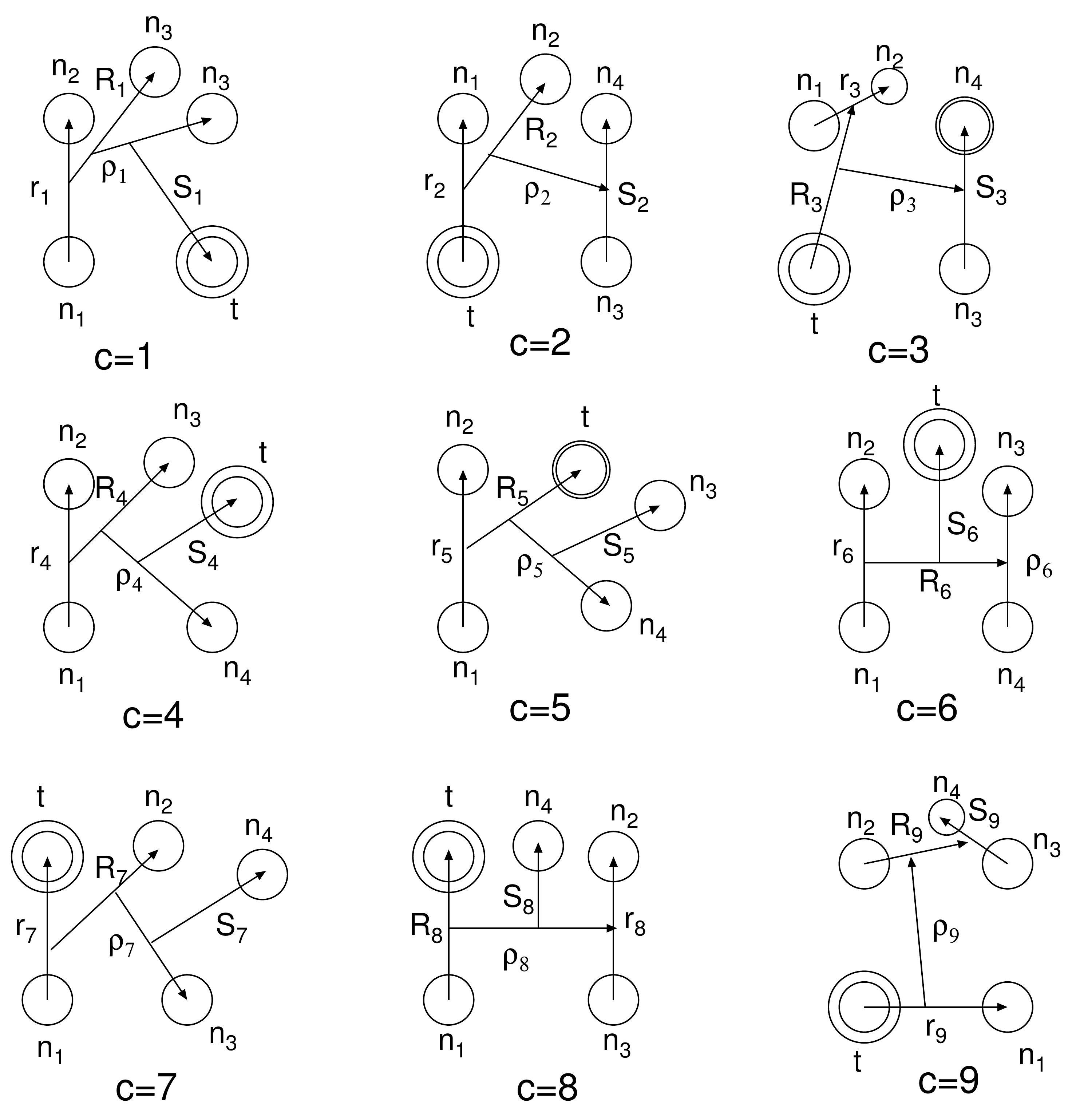}
\caption{Different topologies for the 5-body components and corresponding Jacobi coordinates of the t-n-n-n-n system.}\label{FYC}
\vspace{0.cm}
\end{figure}

We aim to  solve   the Schr\"{o}dinger equation for $^5$H  and $^7$H within the framework of  t-n-n and t-n-n-n-n 3- and 5-body cluster models
respectively,
by means of  the Gaussian Expansion Method (GEM) \cite{GEM_Kamimura_PRA38_1988,HK_GEM_Front_Phys_13_2018}.
It consists in first   splitting the total wave function $\Psi$ in terms of 3  (and 9) components, which correspond to the schematic decomposition
of the 3- (and 5-body) systems in different clusters. These components are
depicted in Figures \ref{FC_5H} and  \ref{FYC}, together  with the corresponding  set of  Jacobi coordinates on which they are naturally expressed.

The total Hamiltonian and the Schr\"{o}dinger equation are given by 
\begin{eqnarray}
 &&  H  =K+\sum_{a,b}V_{a b}   \;+ \sum_{abc}  V_{abc}        \label{eq:hamiltonian}  \\
 && ( H - E ) \, \Psi^{JM}_{TT_z}(^{5,7}{\rm H})  = 0 \label{SCHRO}  \label{H}  
\end{eqnarray}
where $K$ is the kinetic-energy operator, $V_{ab}$ is the interaction between the  constituent particle $a$ and $b$ and $V_{abc}$ the three-body force.

The total wavefunctions of $^5$H and $^7$H are expressed as sum of amplitudes corresponding
to the rearrangement channels  shown in  Figs  \ref{FC_5H} and  \ref{FYC}. In the $LS$ coupling scheme they read, respectively:
\begin{small}
\begin{eqnarray}
\Psi^{JM}_{TT_z}( ^{5}{\rm H})
       \!\!\!&=&\!\!\!  \sum_{c=1}^{3} \;  \sum_{n,N}  \; \sum_{l,L}   \;   \sum_{S,\Sigma,I}       C^{(c)}_{nlNL S\Sigma I}  \cr
      &\times & {\cal A}_N  \Big[       \Phi(t) \big[\chi_{\frac{1}{2}}(t)      \big[ \chi_{\frac{1}{2}}(n_1)      \chi_{\frac{1}{2}}(n_2)     \big]_S \big]_\Sigma   
          \times      \big[ \phi^{(c)}_{nl}({\bf r}_c)         \psi^{(c)}_{NL}({\bf R}_c)\big]_I  \Big]_{JM}   
        \;        \Big[\eta_{\frac{1}{2}}(t)        \big[ \eta_{\frac{1}{2}}(n_1)       \eta_{\frac{1}{2}}(n_2)       \big]_t \Big]_{T,T_z}    \label{Psi_5H}  \\
 \Psi^{JM}_{TT_z}(^{7}{\rm H})
       \!\!\!&=&\!\!\!   \sum_{c=1}^{9} \;      \sum_{n,N}   \; \sum_{l,L,\lambda,\alpha,\beta}   \;    \sum_{S,\Sigma,I,K}    \;   C^{(c)}_{nlNL\lambda\alpha\beta S\Sigma IK}   \;\;  {\cal A}_N   \cr
      &  \times & 
       \Big[
       \Phi(t) \big[\chi_{\frac{1}{2}}(t)
       \big[ \big[ \big[ \chi_{\frac{1}{2}}(n_1)
       \chi_{\frac{1}{2}}(n_2)
       \big]_s  \chi_{\frac{1}{2}}(n_3) \big]_\Sigma 
       \chi_{\frac{1}{2}}(n_4) \big]_{S_4} \big]_S  %\nonumber \\
         \big[ \big[ \big[ \phi^{(c)}_{nl}({\bf r}_c)
         \psi^{(c)}_{NL}({\bf R}_c)\big]_I
        \xi^{(c)}_{\nu\lambda} (\mbox{\boldmath $\rho$}_c)
        \big]_{I_4}   \zeta^{(c)}_{\alpha \beta} ({\bf s}_c) \big]_K \Big]_{JM}\: \nonumber \\
        & \times  &         \big[\eta_{\frac{1}{2}}(t)         \big[ \big[\big[ \eta_{\frac{1}{2}}(n_1)       \eta_{\frac{1}{2}}(n_2)
       \big]_t  \eta_{\frac{1}{2}} (n_3) \big]_\tau \eta_{\frac{1}{2}}(n_4) \big]_{T_4}          \big]_{T,T_z}    \;  ,     \label{Psi_7H}
\end{eqnarray}
\end{small}

$\cal{A}_N$ stands for antisymmetrization operator between  two or four neutrons.
$\chi_{\frac{1}{2}}(t)$ and $\chi_{\frac{1}{2}}(n_i)$ are the
triton and neutron spin functions respectively, and
$\eta_{\frac{1}{2}}(t)$ and $\eta_{\frac{1}{2}}(n_i)$ are  triton and neutrons  isospin functions respectively.
The total isospin (and its projection on z-axis)  $T (T_z)$ are equal 3/2(-3/2) and 5/2(-5/2) for $^5$H and $^7$H nuclei respectively.
In the GEM  the functional form of $\phi_{nlm}({\bf r})$, $\psi_{NLM}({\bf R})$, $\xi^{(c)}_{\nu\lambda\mu}(\mbox{\boldmath $\rho$}_c)$ and $\zeta^{(c)}_{\kappa \alpha \beta} ({\bf s}_c)$
is: 
\begin{eqnarray}
\phi_{nlm}({\bf r})       &=&      r^l \, e^{-(r/r_n)^2}       Y_{lm}({\widehat {\bf r}})  \;  , \nonumber \\
\psi_{NLM}({\bf R})      &=&       R^L \, e^{-(R/R_N)^2}       Y_{LM}({\widehat {\bf R}})  \;  , \nonumber \\
\xi_{\nu\lambda\mu}(\mbox{\boldmath $\rho$}) &=&
\rho^\lambda \, e^{-(\rho/\rho_\nu)^2} Y_{\lambda\mu}({\widehat {\rhovec}})  \; , \nonumber \\
\zeta_{\kappa \alpha \beta} ({\bf s})    &=&  s^\alpha \, e^{-(s/s_\beta)^2}     Y_{\alpha \beta}({\widehat {\bf s}}) \:,
%(2.4)
\end{eqnarray}
where the Gaussian range parameters are chosen  according to geometrical progressions:
\begin{eqnarray}
      r_n   &=&   r_1 a^{n-1} \qquad \enspace      (n=1 - n_{\rm max}) \; , \nonumber\\
      R_N      &=&       R_1 A^{N-1} \quad      (N \! =1 - N_{\rm max}) \; ,    \nonumber\\
     \rho_\nu  &=&      \rho_1 \tilde{a}^{\nu-1} \qquad     (\nu \! =1 - \nu_{\rm max}) \; , \nonumber \\
   s_\beta    &=&    s_1\tilde{A}^{\kappa -1} \qquad     (\kappa =1-\kappa_{\rm max} )\:.
%(2.5)
\end{eqnarray}

The eigenenergy $E$  in Eq. (\ref{SCHRO}) and the coefficients $C^{(c)}$ 
are  determined by {inserting respectively  the expansions (\ref{Psi_5H}) and (\ref{Psi_7H}) in  (\ref{SCHRO})  and applying} the Rayleigh-Ritz variational method.
This results into a generalized eigenvalue equation $AC=\lambda BC$ of typical  dimension d$\sim 60 000$. 
It is worth noticing that the different, not orthogonal,  components of the total wave function  represented  in Fig. \ref{FYC}, are  all crucial
in order to reach a converged solution of the Schr\"{o}dinger equation with  limited numerical resources.

%%%%%%%%%%%%%%%%%%%%%%%%%
\subsection{The stabilization method}\label{SM}

\bigskip
To access the complex energy eigenvalues of a resonant state, we
have used the so-called stabilization (or real scaling) method  \cite{Taylor_AdvChPh28_1970,Simmons_JChPh75_1988}.
It consists in  computing the real eigenvalues of the Hamiltonian matrix representation and studying their (approximate)  "stability"
when varying a scaling parameter (generically denoted $\alpha$) of  the basis sate, for instance, the Gaussian range parameters $\nu_n\to \alpha \nu_n$.

\vspace{0.cm}\begin{figure}[h!]
\centering\includegraphics[width=8.cm]{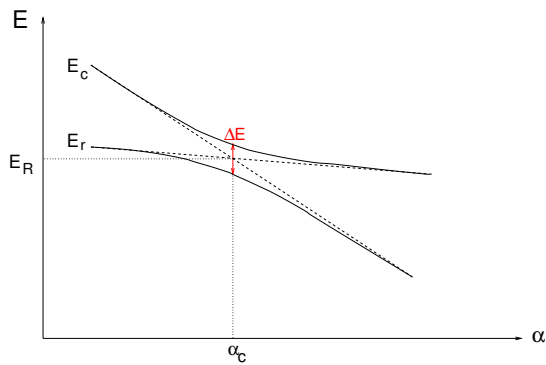}
\vspace{-0.5cm}
 \caption{Example of stabilization graph  as a function of the stabilization parameter $\alpha$}\label{Fig_SG}
\end{figure}

The method is  based in  the observation, illustrated in Figure \ref{Fig_SG},  that when varying $\alpha$, the eigenvalue corresponding to the real part of a resonant state ($E_r$)
has an avoided crossing with the real eigenvalues corresponding to the continuum ($E_c$).
The analysis of this level repulsion near the crossing point, defined by the intersection of the asymptotes to both curves,
provides an estimation of the resonance parameters ($E_R,\Gamma$).
In Figure \ref{Fig_SG}, the  level crossing  between a resonant state ($E_r$) and a continuous state ($E_c$)  moving down, is produced at
 $\alpha=\alpha_c$ and  defines the resonant energy  $E_R$ (dashed horizontal line).
This feature is a signature of a resonance state of the  Hamiltonian with real energy $E_R$ and a width given by \cite{Simmons_JChPh75_1988}
\begin{equation}\label{Gamma_SM}
\Gamma= 2 \Delta E \;  \frac{ \sqrt{\mid S_r \mid \mid S_c\mid}} { \vert S_r - S_c\vert}
\end{equation}
where $\Delta E$  (in red) is the energy difference between the two curves at the crossing, and $S_r$ and $S_c$ are
respectively the slopes of the resonant and continuous levels asymptotes at the crossing.

The stabilization method is closely connected with the Complex Scaling Method (CSM)  \cite{Mo98} but lacks of a sound
mathematical ground, providing only a rough estimation of the resonance width. This method has been recently applied, in
conjunction with the Gaussian Expansion Method, to determine the resonance positions of a doubly heavy tetraquark system \cite{MHHHO_PLB824_2022}.

%%%%%%%%%%%%%%%%%%%%%%%%%%%%%%%%%%%%%%%%
\section{Results}\label{Results}

Our initial effort was  to establish properties of  the $^5$H ground state within our model, i.e., by considering it as  a t-n-n three-body system interacting 
via the t-n potential of eq.(\ref{Vnt})  and Minnesota nn pairwise interaction \cite{NN_Minnesota_1977} in conjunction with the hyperradial 3-body force eq.(\ref{Vtnn}). 
 The solution was obtained, as in Ref. \cite{HOKY_NPA908_2013}, by
combining GEM and CSM, which directly provides the position of the S-matrix pole in the complex energy plate corresponding  to the resonant state.

\begin{figure}[h!]
\centering\includegraphics[width=12.cm]{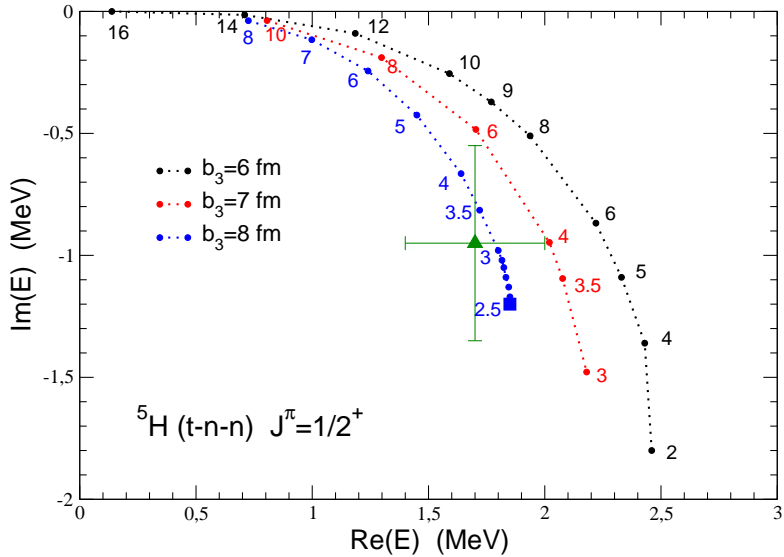} 
\vspace{-0.5cm}\caption{Energy trajectory (E=$E_R$+i $E_I$) of $^5$H ground state in
complex plane, obtained by varying the strength $V_0$ of three-body force (\ref{Vtnn}) for three different values of $b_3$.
The numbers along the curves denote the values of $V_0$ (in MeV).
The filled blue square corresponds to the {\it ab-initio} result of the $^5$H ground state computed in Ref. \cite{5H_LHC_PLB791_2019}; the up-pointing triangle in green
corresponds to the experimental value from \cite{Korsheninnikov_PRL87_2001}.}\label{Fig_E5H_V0} \vspace{0.cm}
\end{figure}

The complex energy   values E=$E_R$+i $E_I$
($E_I\equiv{\Gamma/2}$)   of the $^5$H   J$^{\pi}$=1/2$^+$ ground
state   as a function of the three-body force strength $V_0$,
defined in  eq (\ref{Vtnn}), are displayed in Figure
\ref{Fig_E5H_V0}  in the ($E_R$ ,$E_I$) plane for several values
of the range parameter $b_3$. The numbers along the curves denote
the $V_0$ values in MeV. The {\it ab-initio} value for this state
computed in  \cite{5H_LHC_PLB791_2019} is indicated by a filled
blue  square  (\crule[blue]{0.2cm}{0.2cm}) and the experimental
data from \cite{Korsheninnikov_PRL87_2001} by  a filled green up-pointing triangle
($\color{green}{\bigtriangleup}\color{black}$) with corresponding error bars.

As one can see, the best result is obtained with $b_3$=8 fm and $V_0$=2.5 MeV for which one gets E($^5$H)=1.9 -  1.2 i MeV in full
agreement with our previous {\it ab-initio} theoretical result,
obtained for the realistic NN interaction models
\cite{5H_LHC_PLB791_2019}. This value is quite close to, but different from, the experimental one
\cite{Korsheninnikov_PRL87_2001} as the later one reflects the
footprint of the $^5$H resonant state in a complex nuclear
reaction, which strongly depends on the reaction mechanism but
also on the experimental protocol, while its relation with the S-matrix singularity is only indirect. 

Notice that  reproducing the phenomenology and the {\it ab-initio} results  requires a relatively large  value of the range parameter $b_3$
compared in comparison with Refs. \cite{DGFJ_NPA786_2007,HOKY_NPA908_2013}. 
This is related to the choice of our n-t potential, with the particular functional form of  eq. (\ref{Vnt}). 
The presence of strongly repulsive S-wave channels, manifested by the $\phi_o$ term with parameter  $a_0\approx$ 0.20 fm$^{-2}$,  
implies the exclusion of a large spatial region to project out the symmetry forbidden states in $^4$H. 
Any attractive contribution, aimed  to account for the spin-dependent effects in the n-t S-waves,  as well as to 
describe the energy dependence of the n-t phase shifts, will require an even larger interaction range to have a significant effect. 
Furthermore, any attempt to reproduce the  $^5$H resonance position also favored large interaction ranges. 
This  is an obvious consequence  of simulating the Pauli  repulsion between the valence neutrons and those  of the triton core. 
As the n-n interaction is resonant in $^1$S$_0$ state,  dineutron correlations take place at rather large distances, and this generates also a rather large exclusion region.

\begin{figure}[h!]
%\centering\includegraphics[width=8.cm]{scal-h7.pdf} 
\centering\includegraphics[width=7.cm]{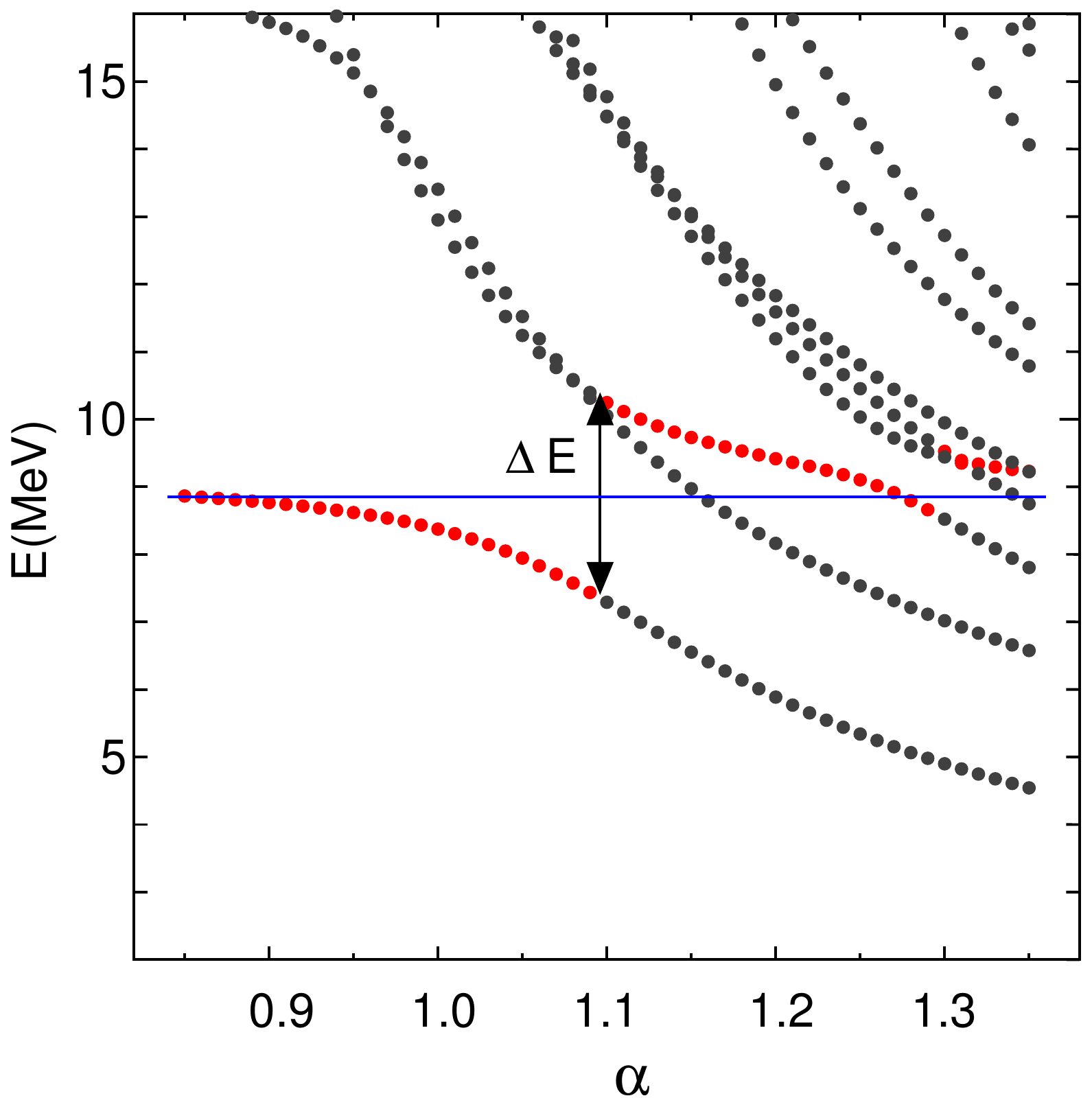} 
\vspace{-0.3cm}\caption{$^7$H stabilization graph: lower eigenvalues of the $^7$H 
hamiltonian as a function of the stabilization parameter $\alpha$ corresponding
to the three-body force parameters $b_3$=8 fm and $V_0$=3 MeV in (\ref{Vtnn}).}\label{Fig_7H_SG} \vspace{0.cm}
\end{figure}

Having fixed the parameters of the t-n and t-n-n interactions, we have estimated the complex energy of $^7$H ground state by using the stabilization method, briefly described in section (\ref{SM}). 
An illustrative example of  results provided by this approach in the case of $^7$H, is displayed in Figure \ref{Fig_7H_SG}. 
It reveals the stabilization graph $E(\alpha)$ -- that is the lower values of the $^7$H Hamiltonian spectra  as a
function of the stabilization parameter $\alpha$ -- corresponding to the 3-body parameters $b_3$=8 fm and $V_0$=3 MeV.

In the last figure a level repulsion manifests at $\alpha\approx$1.1  in the energy range
E= 7-10 MeV  with $\Delta E\approx$ 3 MeV, indicated by a vertical double arrow. This is a signature of a resonant state at $E_R\approx$  8.8 MeV  (denoted by an
horizontal blue line) and an estimated width $\Gamma=2E_I\approx$ 3.1 MeV given by eq. (\ref{Gamma_SM}).

Following this approach, we have  computed the $^7$H ground state
energy as a function of the t-n-n three-body force strength parameter $V_0$. The range of  $V_{tnn}$ was fixed to $b_3$=8 fm
and we have varied its strength $V_0$  from $V_0=7$ MeV until it reaches the value $V_0=2.5$, closely reproducing position of the
$^5$H resonant state established in our previous ab-initio calculations~\cite{5H_LHC_PLB791_2019}.

\begin{figure}[h!]
\centering\includegraphics[width=12.cm]{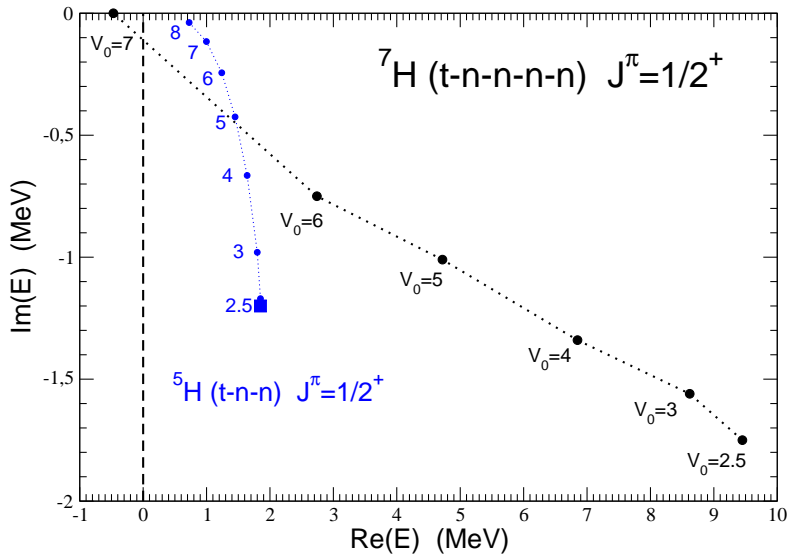}
\vspace{-0.5cm}\caption{In black, the $^7$H ground state complex energy   trajectory (E=$E_R$+i $E_I$) as a function of the strength $V_0$ of three-body force (\ref{Vtnn}).
Results ware obtained with fixed value $b_3$=8 fm, reproducing the ab-initio $^5$H ground state.
The results for $^5$H (corresponding to Fig \ref{Fig_E5H_V0}) are displayed in blue for comparison,
and the  filled blue square corresponds to the {\it ab-initio} result of the $^5$H ground state
computed in Ref. \cite{5H_LHC_PLB791_2019}.}\label{Fig_E7H_V0}
\vspace{0.cm}
\end{figure}

The final results are summarized in Figure \ref{Fig_E7H_V0}.
The complex energies of $^7$H are indicated by black dots joined by black dashed line. 
For comparison we have also included in the same figure (blue dots joined by blue dashed line) the results of $^5$H, already described in Figure \ref{Fig_E5H_V0}.

For $V_0$=7 MeV, the $^7$H appears to be bound  with respect to
the t+4n threshold by E$\approx$-0.5 MeV. When decreasing $V_0$
this bound state turns into a resonance. When the strength of
t-n-n force is further reduced to its realistic value $V_0$=2.5 --
reproducing position of the $^5$H  -- the $^7$H resonance is found
at $E_R\approx$ 9.5  MeV with an estimated width
$\Gamma=2E_I\approx$ 3.5 MeV. If we increase the value $V_0=3$
MeV, producing $^5$H ground state at the lowest point compatible
with the experiment, the $^7$H resonance is found at $E_R=8.5$ MeV
and $\Gamma=3$ MeV. These finding conflicts with a presence of a
narrow $^7$H resonant state in the vicinity of t+4n threshold. We
still find a signature of a broad state at relatively high energy
sufficient to break triton core in single particles. This high
energy domain is beyond the range of validity of our model, which
supposes existence of a solid tritium core. One expects that the
widths of $^7$H state should become even larger in a more
realistic model, where tritium core is allowed to break.

As one can see our results are in  no way compatible with the existence of a narrow $^7$H resonant state
as being claimed in some experimental works, mainly \cite{7H_GANIL_Fortier_AIPCP912_2007,7H_GANIL_Baumel_ISPUN0_2007,7H_GANIL_PRL99_2007,7H_RIKEN_PRC81_2010,7H_Dubna_PRL124_2020}.
Although all these experiments suffer from a very poor statistics, there is a huge gap between these two views.
The only possible link between our results and the experimental claims is to identify our values with what the authors of Ref.  \cite{7H_Dubna_PRL124_2020} claim to be
the first excited state of $^7$H with $E^*$=6.5(5) and $\Gamma$=2.0(5).

In relation with the previous theoretical works, our conclusions are in the same line than those of Ref.
\cite{Aoyama_Itagaki_NPA738_2004} who obtain E($^7$H)$\sim$ 7 MeV.
Our findings are however  in a sharp contrast with the recent work  \cite{Li2_Michel_Zuo_PRC104L06_2021}. These authors provide an
energy of the ground $^7$H which varies from 2-3 MeV, depending on
the NN interaction considered, and in all cases a very narrow
width  $\Gamma \approx0.1$ MeV (see Fig 4 of this reference, with a factor 10 enhancement in the $^7$H widths). 

It is worth emphasizing that our differences with the results of \cite{Li2_Michel_Zuo_PRC104L06_2021} come already from the
very input of both approaches. Indeed, in a former work \cite{Li2_Michel_Zuo_PRCH024319}, the same authors using the same method   found very narrow $^4$H resonant states (see Fig 1), 
placing them at $E_R\approx$1.7- 0.9 i MeV ,  in agreement with some  R-matrix analysis \cite{Gurov_EPJA24_2005} . 
However, it is well established that experimental R-matrix results and the S-matrix pole positions may
differ significantly for broad resonant states, which is in
particular true for $^4$H case as stated in \cite{Tiley_A_4_NPA451_1992,DGFJ_NPA786_2007,5H_LHC_PLB791_2019}. As a consequence of this original choice,
\cite{Li2_Michel_Zuo_PRC104L06_2021} obtained also an extremely narrow $^5$H resonance (see Fig 4), which is not supported neither by theory nor by the experiment.  

In our opinion, there seems to exist an intrinsic limitation in the GSM used in
\cite{Li2_Michel_Zuo_PRC104L06_2021}, which makes this method to systematically underestimate the widths of  resonant states. 
It  is due to the fact that GSM employs only single
particle basis, which are  not well adapted to describe collective cluster channels and thus limits the decay possibilities of a resonant states. 
In  the same way, if we would have restricted  the variation of the basis parameters in the stabilization method described in Section \ref{SM}, for instance by  just limiting 
to few selected components of Figure \ref{FYC}, we  would have obtained a significant  reduction of the estimated  $^7$H width.

For the sake of completeness, we have also computed the resonant ground state of $^6$H, considered as a t-n-n-n four-body system
By using the same nn, t-n, and t-n-n  interactions than those used to obtain the $^5$H and  $^7$H,  in particular $V_0$=2.5 and $b_3=$8 fm,
we estimate for $^6$H ground state the resonance parameters $E_R$=10 MeV and $\Gamma=$4 MeV.
Both, energy and width, are sligthly larger  than those of $^7$H, though compatible.
 
\begin{figure}[h!]
\centering\includegraphics[width=12.cm]{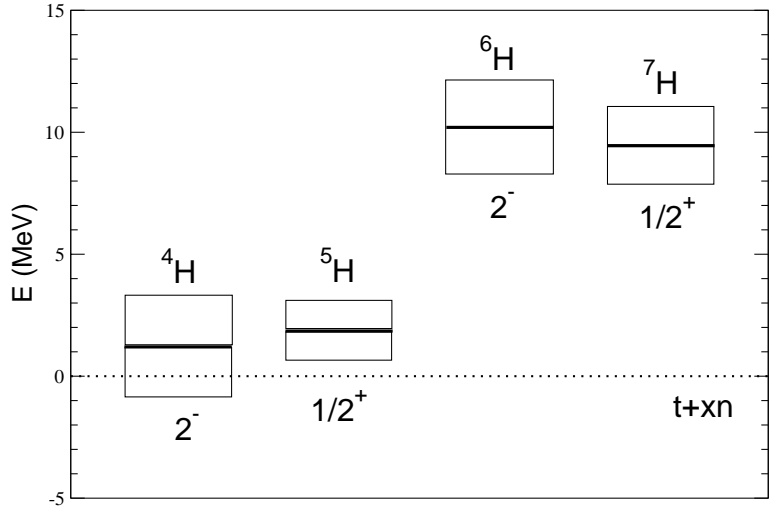}
\vspace{-0.5cm}
\caption{Predictions for the  ground states of heavier H-isotopes. 
$^7$H ground state ($E_R$=9.5 MeV and $\Gamma$=3.5 MeV)  is obtained in the t-n-n-n-n cluster approach with the  parameters of
the t-n-n interaction (b=8 fm  $V_0$=2.5 MeV)  adjusted  to reproduce  the $^5$H ground state. $^4$H ($2^-$) and $^5$H  values
are those provided by {\it ab-initio} calculations with realistic NN interactions and are taken from
\cite{5H_LHC_PLB791_2019}.}\label{Spectrum_4H_5H_7H} \vspace{0.cm}
\end{figure}

\bigskip
In summary,  our theoretical analysis of the heavier well
identified H-isotopes  $^4$H, $^5$H, $^6$H and $^7$H can be described in
Figure \ref{Spectrum_4H_5H_7H}. As it is customary in the
experimental tables, it represents the resonance energies $E_R$ of
their ground states relative to the corresponding t+$x$n
threshold, with an overimposed square denoting the total width $\Gamma$ (all units are in MeV).

The results of $^6$H ($E_R$=10 MeV, $\Gamma$=4 MeV)  and $^7$H ($E_R$=9.5 MeV, $\Gamma$=3.5 MeV) are those obtained
in the present work, within  the t-n-n-n-n cluster approximation and $V_{tnn}$ parameters $b_3$=8 fm  and $V_0$=2.5 MeV.
{They are compatible with the experimental value  $E_{gs}$=7.3$\pm$1 MeV, $\Gamma_{gs}$=5.8 $\pm$ 2 MeV
obtained by \cite{Gurov_EPJA24_2005} in the  $^{11}$B($\pi^-$,p$^4$He)$^6$H reaction. They indicate that the big jump in $E_R$  already manifests in $^6$H.}

The values  for $^4$H  and $^5$H correspond  respectively  to  $2^-$  ($E_R$=1.2 MeV, $\Gamma$=4 MeV)
and $1/2^+$  ($E_R$=1.85 MeV, $\Gamma$=2.4 MeV),
  both supposed to be the experimental ground states,  as they are  provided by the ab initio calculations and N3LO
interaction in our previous work \cite{5H_LHC_PLB791_2019} (see
Table I and II). For $^5$H, the energy  fully coincides with the
value  of  the t-n-n cluster approximation displayed in  Figs
\ref{Fig_E5H_V0} (blue squared corresponding to $b_3$=8 fm and $V_0$=2.5 MeV). For $^4$H, it is slightly different from the
$J^{\pi}$=0$^-$,1$^-$,2$^-$ degenerate value of the two-body ground state ($E_R$=1.3 MeV, $\Gamma$=3.8 MeV) provided by the n-t potential (\ref{Vnt}).

The tendency to stability observed between $^4$H and $^5$H, pointed out in our
previous work \cite{5H_LHC_PLB791_2019}, is totally
reversed by the heavier isotopes. Both,  $^6$H  and $^7$H ground states are at relatively high energies ($E_R\sim$ 9 MeV) and are broad ($\Gamma \sim$ 4 MeV)
resonant states, which suggest  the end point of the H isotopic chain  rather than a stability region opened by the neutron richer isotopes.

In view of the strong contradiction with previous experimental and theoretical works,
it would be highly desirable that new high statistic experiments and independent calculations could bring some light into this disarming situation.

%%%%%%%%%%%%%%%%%%%%%%%%%%%%
\section{Conclusion}\label{Conclusion}

We have obtained   the complex energy  position of  the $^7$H resonant ground state  considered as a  5-body t+n+n+n+n system.

To this aim, we first constructed a n-t local potential whose
parameters were adjusted to reproduce the low energy neutron
scattering on triton phase shifts. An additional 3-body force was
needed to reproduce the $^5$H ground state in the t+n+n approximation.

The solution of the 5-body  Schr\"{o}dinger equation has been
obtained by means  of the variational Gaussian expansion approach
and the resonance parameters have been estimated  by using the stabilization (or real scaling) method.

Our estimated parameters of $^7$H ground state are $E_R$=9.5 MeV
and $\Gamma$=3.5 MeV, that is quite a broad state when compared
with the lighter H isotopes ($^4$H and $^5$H). Our findings are in
sharp contrast with some experimental values
\cite{7H_RIKEN_Korsheninnikov_PRL90_2003,7H_GANIL_PRL99_2007,7H_RIKEN_PRC81_2010,7H_Dubna_PRL124_2020}
as well as with the theoretical results of \cite{Li2_Michel_Zuo_PRC104L06_2021}.

By using the same interaction parameters than for $^7$H, a similar calculation was performed to describe the ground state of the $^6$H isotope,  considered as a $^3$H+3n four-body system.
The ground state was identified to have $J=2^-$  and is situated at $E_R\approx$10 MeV and $\Gamma\approx$4 MeV.

Even within the limits of our calculation,  we must exclude the presence of a narrow resonant state in $^7$H. 
The comparison with $^4$H and $^5$H makes unlikely  that the H isotopic chain could be continued, at least an experimentally observed state, beyond $^7$H.

\section*{Aknowledgements}

We are sincerely grateful  to  V. Lapoux and F. M. Marqu\'es for enlightening discussions concerning the   experimental results.
We were granted access to the HPC resources of TGCC/IDRIS under the allocation A0110506006 by GENCI (Grand Equipement National de Calcul Intensif).
This work was supported by french IN2P3 for a theory project "Neutron-rich light unstable nuclei",
by the japanese  Grant-in-Aid for Scientific Research on Innovative Areas (No.18H05407)
and by the Pioneering research project  'RIKEN Evolution of Matter in the Universe Program'.
{It was completed during the program Living Near Unitarity at Kavli Institute for Theoretical Physics,  University of Santa Barbara (California).
We thank the organizers and the staff members of this Institute for their kind invitation and financial support, from the National Science Foundation  Grant No. NSF PHY-1748958.}

%%%%%%%%%%%%%%%%%%%%%%%%%%%%%%%%%%%%%%%%%%%%%%%%%%%%%%%

\end{document}